\documentstyle[preprint,floats,pra,aps]{revtex}
\tightenlines

\def\beq{\begin{equation}}
\def\eeq{\end{equation}}
\def\bea{\begin{eqnarray}}
\def\eea{\end{eqnarray}}
\def\nn{\nonumber}
\def\ba{\begin{array}}
\def\ea{\end{array}}   
\def\i{{\rm i}}
\begin{document} \draft

\title{\LARGE\bf Quantum-like approaches to the beam halo problem
\footnote{
{\em To appear in}: Proceedings of the 6th International
Conference on Squeezed States and Uncertainty Relations, (ICSSUR'99),
24-29 May 1999, Napoli, ITALY, (NASA Conference Publication Series).}}

\author{Sameen Ahmed KHAN}
\address{
Dipartimento di Fisica Galileo Galilei  
Universit\`{a} di Padova \\
Istituto Nazionale di Fisica Nucleare~(INFN) Sezione di Padova \\
Via Marzolo 8 Padova 35131 ITALY \\
E-mail: khan@pd.infn.it, ~~~ http://www.pd.infn.it/$\sim$khan/}

\author{Modesto PUSTERLA} 
\address{
Dipartimento di Fisica Galileo Galilei  
Universit\`{a} di Padova \\
Istituto Nazionale di Fisica Nucleare~(INFN) Sezione di Padova \\
Via Marzolo 8 Padova 35131 ITALY \\
E-mail: pusterla@pd.infn.it, ~~~ http://www.pd.infn.it/$\sim$pusterla/}

\maketitle

\begin{abstract}
An interpretation of the the ``halo problem'' in accelerators based on
quantum-like diffraction is given. Comparison between this approach 
and the others based on classical mechanics equations is discussed
\end{abstract}


\vspace{8mm}

\section{Introduction}
Recently the description of the dynamical evolution of high density 
beams by using the collective models, has become more and more popular.
A way of developing this point of view is the quantum-like 
approach~\cite{Fedele} where one considers a time-dependent 
Schr\"{o}dinger equation, in both the usual linear and the less usual 
nonlinear forms, as a fluid equation for the whole beam. In this case 
the squared modulus of the wave function~(named beam wave function) gives
the distribution function of the particles in space at a certain 
time~\cite{PAC}. The Schr\"{o}dinger equation may be taken in one or more
spacial dimensions according to the particular physical problem; 
furthermore the fluid becomes a Madelung fluid if one choses the equation
in its usual linear version.

Although the validity of the model relies only on experiments and in
particular on new predictions which must be verified experimentally,
we like to invoke here a theoretical argument that could justify the
Schr\"{o}dinger quantum-like approach we are going to apply. Let us
think of particles in motion within a bunch in such a way that the 
single particle moves under an average force field due to the presence 
of others and collides with the neighbouring ones in a complicated 
manner. It is obviously impossible to follow and predict all the forces
deterministically.  We then face a situation where the classical motion 
determined by the force-field is perturbed continuously by a random term,
and we have a connection with a stochastic process. If one simply assumes
that the process is Markovian and Brownian, one obtains following 
Nelson~\cite{Nelson}, a modification of the classical equations
of motion that can be synthesized by a linear Schr\"{o}dinger equation 
which depends on a physical parameter having the dimension of 
action~\cite{Guerra}. Usual wave quantum mechanics follows if this 
parameter is chosen as the Planck's constant~$\hbar$, whereas the 
quantum-like theory of beams in the linearized version is obtained if one
choses the normalized emittance~$\epsilon$~\cite{Fedele}. In both cases,
quantum particle and beam respectively, the evolution of the system is 
expressed in terms of a continuous field~$\psi$ defining the so called 
Madelung fluid. We may notice that the normalized emittance~$\epsilon$
having the dimension of action is the natural choice for the parameter in 
the quantum-like theory that corresponds to Planck's constant~$\hbar$ in 
the quantum theory, because it reproduces the corresponding area in the 
phase-space of the particle.

We here point out that, after linearizing the Schr\"{o}dinger-like
equation, for beams in an accelerator one can use the whole apparatus
of quantum mechanics, with a new interpretation of the basic parameters
(for instance the Planck's constant $\hbar \longrightarrow \epsilon$ 
where $\epsilon$ is the normalized beam emittance) and introduce the 
propagator $K \left( x_f , t_f | x_i , t_i \right)$ of the Feynman theory 
for both longitudinal and transversal motion. A procedure of this sort 
seems particularly effective for a global description of several phenomena 
such as intrabeam scattering, space-charge, particle focusing, that cannot
be treated easily in detail by ``classical mechanics'' and are considered 
to be the main cause for the creation of the {\em Halo} around the beam 
line with consequent losses of particles.

Let us indeed consider the Schr\"{o}dinger like equation for the beam
wave function
\bea
\i \epsilon \partial _t \psi 
= - \frac{\epsilon^2}{2 m} \partial_x ^2 \psi + U \left( x , t \right) \psi
\label{schroedinger-like}
\eea
in the linearized case $U \left (x , t \right)$ does not depend on the 
density $\left| \psi \right|^2$. $\epsilon$ here is the normalized
transversal beam emittance defined as follows:
\bea
\epsilon = m_0 c \gamma \beta \tilde{\epsilon}\,,
\label{epsilon}
\eea
$\tilde{\epsilon}$ being the emittance usually considered, where as
we may introduce the analog of the de Broglie wavelength as
$\lambda = {\epsilon}/{p}$. We now focus our attention on the one 
dimensional transversal motion along the $x$-axis of the beam particles 
belonging to a single bunch and assume a Gaussian transversal profile 
for a particles injected in to a circular machine. We describe all the 
interactions mentioned above, that cannot be treated in detail, as 
diffraction effects by a phenomenological boundary defined by a slit,
in each segment of the particle trajectory. This condition should be 
applied to both beam wave function and its corresponding beam
propagator $K$. The result of such a procedure is a multiple integral
that determines the actual propagator between the initial and final states
in terms of the space-time intervals due to the intermediate segments.
\bea
K \left(x + x_0 , T + \tau | x' , 0 \right) 
& = &
\int_{- b}^{+ b}
K \left(x + x_0 , \tau | x_0 + y_n , T + (n - 1) \tau ' \right) \nn \\
& & \quad \times 
K \left(x + y_n , T + (n - 1) \tau ' | 
x_0 + y_{n - 1} , T + (n - 2) \tau ' \right) \nn \\
& & \qquad \qquad \qquad \times \cdots 
K \left(x + y_1 , T | x' , 0 \right) d y_1 d y_2 \cdots d y_n 
\label{integral}
\eea
where $\tau = n \tau '$ is the total time of revolutions $T$ is the
time necessary to insert the bunch (practically the time between two
successive bunches) and $(-b , +b)$ the space interval defining the
boundary conditions.
Obviously $b$ and $T$ are phenomenological parameters which vary from
a machine to another and must also be correlated with the geometry of the
vacuum tube where the particles circulate.

At this point we may consider two possible approximations for 
$K \left( n | n - 1 \right) \equiv 
K \left( x_0 + y_n , T + (n - 1) \tau ' | 
x_0 + y_{n - 1} + (n - 2) \tau ' \right)$:

\begin{enumerate}

\item
We substitute it with the free particle $K_0$ assuming that in the
$\tau '$ interval $(\tau ' \ll \tau)$ the motion is practically a free
particle motion between the boundaries $( -b , + b )$.

\item
We substitute it with the harmonic oscillator 
$K_{\omega} \left( n | n -1 \right)$
considering the harmonic motion of the betatronic oscillations with
frequency $\omega/{2 \pi}$

\end{enumerate}

\section{Free Particle Case}
We may notice that the convolution property~(\ref{integral}) of the
Feynman propagator allows us to substitute the multiple integral
(that becomes a functional integral for $n \longrightarrow \infty$ and 
$\tau ' \longrightarrow 0$) with the single integral
\bea
K \left( x + x_0 , T + \tau | x' , 0 \right) 
= \int_{- b}^{+ b} dy
K \left( x + x_0 , T + \tau | x_0 + y , T \right) 
K \left( x_0 + y , T | x' , 0 \right) dy
\label{single}
\eea

In this note we mainly discuss the case~1. and obtain from 
equation~(\ref{single}) after introducing the Gaussian slit
$\exp{\left[- \frac{y^2}{2 b^2}\right]}$ instead of the 
segment $\left( - b , + b \right)$ we obtain from
\bea
& & K \left(x + x_0 , T + \tau | x' , 0 \right) \nn \\
& & = 
\int_{-\infty}^{+\infty}  dy
\exp{\left[-\frac{y^2}{2 b^2} \right]} 
\left\{\frac{2 \pi \i \hbar \tau}{m} 
\frac{2 \pi \i \hbar T}{m} \right\}^{- \frac{1}{2}}
\exp{\left[\frac{\i m}{2 \hbar \tau} (x - y)^2\right]} 
\exp{\left[\frac{\i m}{2 \hbar T} (x_0 + y - x')^2\right]} \nn \\
& & = 
\sqrt{\frac{m}{2 \pi \i \hbar}}
\left(T + \tau + T \tau \frac{\i \hbar}{m b^2} \right)^{-\frac{1}{2}} 
\exp
\left[
\frac{\i m}{ 2 \hbar} \left(v_0^2 T + \frac{x^2}{\tau} \right)
+
\frac{\left(m^2/{2 \hbar^2 \tau^2}\right) \left(x - v_0 \tau \right)^2}
{\frac{\i m}{\hbar} \left(\frac{1}{T} + \frac{1}{\tau} \right) 
- \frac{1}{b^2}}
\right]
\label{exp}
\eea
where $v_0 = \frac{x_0 - x'}{T}$ and $x_0$is the initial central point 
of the beam at injection and can be chosen as the origin ($x_0 = 0$) of
the transverse motion of the reference trajectory in the test particle
reference frame. {\bf Where as $\hbar$ must be interpreted as the 
normalized beam emittance in the quantum-like approach}.

With an initial Gaussian profile (at $t = 0$), the beam wave function 
(normalized to 1) is
\bea
f (x) = \left\{ \frac{\alpha}{\pi} \right\}^{\frac{1}{4}}
\exp{\left[- \frac{\alpha}{2} x'^2 \right]}
\eea
r.m.s of the transverse beam and the final beam wave function is:
\bea
\phi (x) 
= 
\int_{- \infty}^{+ \infty} d x'
\left(\frac{\alpha}{\pi} \right)^{\frac{1}{4}}
e^{\left[- \frac{\alpha}{2} x'^2\right]}
K \left(x, T + \tau ; x', 0\right) 
= B \exp{\left[C x^2 \right]}
\eea
with 
\bea
B & = &
\sqrt{\frac{m}{2 \pi \i \hbar}}
\left\{T + \tau + T \tau \frac{\i \hbar}{m b^2}\right\}^{- \frac{1}{2}}
\left\{\frac{\alpha}{\pi}\right\}^{\frac{1}{4}} 
\sqrt{
\frac{\pi}{
\left(
\frac{\alpha}{2} 
- \frac{\i m}{2 \hbar T} 
- \frac{{m^2}/{2 \hbar^2 T^2}}{
\frac{\i m}{\hbar}\left(\frac{1}{T} + \frac{1}{\tau}\right)
- \frac{1}{b^2}}
\right)
}} \nn \\
C & = &
\frac{\i m}{2 \hbar \tau}
+
\frac{{m^2}/{2 \hbar^2 T^2}}{
\frac{\i m}{\hbar}\left(\frac{1}{T} + \frac{1}{\tau}\right)
- \frac{1}{b^2}} 
+
\frac{
\frac{\tau^2}{T^2}
\left\{
\frac{{m^2}/{2 \hbar^2 T^2}}{
\frac{\i m}{\hbar}\left(\frac{1}{T} + \frac{1}{\tau}\right)
- \frac{1}{b^2}} 
\right\}^2}
{
\left(
\frac{\alpha}{2} 
- \frac{\i m}{2 \hbar T} 
- \frac{{m^2}/{2 \hbar^2 T^2}}{
\frac{\i m}{\hbar}\left(\frac{1}{T} + \frac{1}{\tau}\right)
- \frac{1}{b^2}}
\right)
}
\label{BC}
\eea

The final local distribution of the beam that undergoes the diffraction is
therefore 
\bea
\rho (x) = \left| \phi (x) \right|^2 
= B B^{*} \exp{\left[ - \tilde{\alpha} x^2 \right]}
\eea
where $\tilde{\alpha} = - (C + C^{*})$ and the total probability per 
particle is given by
\bea
P = \int_{- \infty} ^{+ \infty} d x \rho ( x ) 
= B B^{*} \sqrt{\frac{\pi}{\tilde{\alpha}}} 
\approx 
\frac{1}{\sqrt{\alpha}} \frac{m b}{\hbar T}
\label{probability}
\eea
One may notice that the probability $P$ has the same order of magnitude
of the one computed in~\cite{Feynman} if $\frac{1}{\sqrt{\alpha}}$ is 
of the order of $b$.

\section{OSCILLATOR CASE}
Similarly we may consider the harmonic oscillator case 
(betatronic oscillations) to compute the diffraction probability of the 
single particle from the beam wave function and evaluate the probability 
of beam losses per particle. The propagator 
$K_{\omega} \left( x , T + \tau | y , T \right)$
in the later case is:
\bea
& & K \left(x , T + \tau | x' , 0 \right) \nn \\
& & =
\int_{- \infty}^{+ \infty}  dy
\exp{ \left[- \frac{y^2}{2 b^2} \right]} 
K_{\omega} \left(x , T + \tau | y , T \right) 
K_{\omega} \left(y , T | x' , 0 \right)  \nn \\
& & =
\int_{- \infty}^{+ \infty}  dy
\exp{\left[- \frac{y^2}{2 b^2} \right]}
\left\{\frac{m \omega}{2 \pi \i \hbar \sin (\omega \tau)}\right\}^{\frac{1}{2}}
\exp \left[
\frac{\i m \omega}{2 \hbar \sin (\omega \tau)}
\left\{ \left(x^2 + y^2\right) \cos \omega \tau - 2 x y \right\}
\right] \nn \\
& & \qquad \qquad \qquad \qquad \quad \times
\left\{\frac{m \omega}{2 \pi \i \hbar \sin (\omega T)}\right\}^{\frac{1}{2}}
\exp \left[
\frac{\i m \omega}{2 \hbar \sin (\omega T)}
\left\{\left(y^2 + {x'}^2\right) \cos \omega T - 2 x' y \right\}
\right] \nn \\
& & =
\left\{\frac{1}{2 \pi} \tilde{C} \right\}^{\frac{1}{2}}
\exp
\left[\tilde{A} x^2 + \tilde{B} {x'}^2 + \tilde{C} x x' \right]
\label{betatron}
\eea
where
\bea
\tilde{A} & = &
\i \frac{m \omega}{2 \hbar} 
\frac{\cos\left(\omega \tau\right)}{\sin\left(\omega \tau\right)}
-
\left(\frac{m \omega}{2 \hbar}\right)^2
\frac{1}{\sin^{2} \left(\omega \tau\right)}
\frac{1}{D}\,, \qquad
\tilde{B} =
\i \frac{m \omega}{2 \hbar} 
\frac{\cos\left(\omega T\right)}{\sin\left(\omega T\right)}
-
\left(\frac{m \omega}{2 \hbar}\right)^2
\frac{1}{\sin^{2} \left(\omega T\right)}
\frac{1}{D} \nn \\
\tilde{C} & = & 
-
\left(\frac{m \omega}{2 \hbar}\right)^2
\frac{2}{\sin\left(\omega \tau \right) \sin\left(\omega T\right)}
\frac{1}{D}\,, \quad \quad \qquad 
D =
\frac{1}{2 b^2} 
-
\i \frac{m \omega}{2 \hbar} 
\left(
\frac{\cos\left(\omega \tau\right)}{\sin\left(\omega \tau\right)}
+
\frac{\cos\left(\omega T\right)}{\sin\left(\omega T\right)}
\right)
\eea

\bea
\phi_{\omega} (x) 
= 
\int_{- \infty}^{+ \infty} d x'
\left(\frac{\alpha}{\pi} \right)^{\frac{1}{4}}
\exp{\left[- \frac{\alpha}{2} x'^2\right]}
K_{\omega} \left(x, T + \tau ; x', 0\right) 
= 
N \exp{\left[M x^2 \right]}
\eea
where
\bea
N = 
\left(\frac{\alpha}{\pi}\right)^{\frac{1}{4}}
\left\{
\frac{\tilde{C}}{\left(\alpha - 2 \tilde{B}\right)} 
\right\}^{\frac{1}{2}}\,, \qquad \qquad 
M =
\tilde{A}
+
\frac{\tilde{C}^2}{2 \left(\alpha - 2 \tilde{B}\right)} 
\eea
%
\bea
\rho_{\omega} (x) = \left| \phi_{\omega} (x) \right|^2 
= N^{*} N \exp{\left[ - \left(M^{*} + M \right) x^2 \right]}
\eea
%

\bea
P_{\omega} & = & \int_{- \infty} ^{+ \infty} d x \rho ( x ) 
= N^{*} N \sqrt{\frac{\pi}{\left(M^{*} + M\right)}}
\approx 
\frac{1}{\sqrt{\alpha}} \frac{m b}{\hbar} 
\frac{\omega}{\sin \left(\omega T\right)}
\label{probability-w}
\eea
From the approximate formulae~\ref{probability} and~\ref{probability-w} 
we notice that the parameter $\tau$ does not play a significant role in 
the calculation of the probabilities. We gave it a value $\tau = 1$ sec.,
considering about $10^{6}$ revolutions in $LHC$ and $HIDIF$ storage rings.

\section{PRELIMINARY ESTIMATES}

\begin{center}

{\bf TABLE-I: Free Particle Case}  \\
\medskip

\begin{tabular}{llll}
{\bf Parameters} & {\bf LHC} & {\bf HIDIF} \\
Normalized Transverse Emittance ~~~~~~~~~  
& $3.75$ mm mrad ~~~~~~~~~~~~~~~~~ & $13.5$ mm mrad \\
Total Energy, $E$  & $450$ GeV & $5$ Gev  \\
$T$ & $25$ nano sec. & $100$ nano sec. \\
$b$ & $1.2$ mm & $1.0$ mm \\
$P$ & $3.39 \times 10^{-5}$ & $2.37 \times 10^{-3}$ \\
\end{tabular}
 
\end{center}

\bigskip

\begin{center}

{\bf TABLE-II: Oscillator Case} \\
\medskip

\begin{tabular}{llll}
{\bf Parameters} & {\bf LHC} & {\bf HIDIF} \\
Normalized Transverse Emittance ~~~~~~~~~  
& $3.75$ mm mrad ~~~~~~~~~~~~~~~~~~ & $13.5$ mm mrad \\
Total Energy, $E$  & $450$ GeV & $5$ Gev  \\
$T$ & $25$ nano sec. & $100$ nano sec. \\
$b$ & $1.2$ mm & $1.0$ mm \\
$\omega$ & $4.47 \times 10^{6}$ Hz & $1.12 \times 10^{7}$ Hz \\
$P_{\omega}$ & $3.44 \times 10^{-5}$ & $2.96 \times 10^{-3}$ \\
\end{tabular}
 
\end{center}

\section{CONCLUSION}
The parameters entering into the probability formulae are very few and
must be looked at as purely phenomenological. To be more specific,
$b$, $\tau$, and $T$~($b$ in particular) hide several fundamental
processes that may be present in the beam bunches and that play a
deterministic role in the creation of the {\bf halo} such as intrabeam 
scattering beamstrahlung, space-charge effects and imperfections in 
the magnets of the lattice that cause nonlinear perturbative effects.

The fact that such a small amount parameters take into account many
physical processes is a nice feature of the quantum-like diffraction
approach. However a way of connecting this method with the physical 
processes mentioned above as well as with the nonlinear dynamical 
classical theory is mandatory at this point.

Another interesting feature of the parameters used is that their 
numerical values are very reasonable because they are within the ranges.
One expects: $T$ may be considered as the average time interval between 
the two successive injection, $\tau$ the time interval between two
successive diffractions~($\tau = n \tau'$ is the total time of revolutions)
and $2 b$ the phenomenological diffraction slit width. We recall that
in the usual optics diffraction through a slit  is also a macroscopic means 
of dealing with many complicated physical effects such as scattering of
light, electrons etc., at the atomic level.

The two relevant concluding remarks are the following:

\begin{enumerate}

\item
The probability calculated from the free and the harmonic oscillator 
propagators~(both in the transversal motion of the particles) appear
very close for the two different circular systems such as $LHC$ and 
$HIDIF$ rings.

\item
The $HIDIF$ scenario, as expected has a total loss of beam power which
is at least $10^3$ times higher than $LHC$.

\end{enumerate}

These preliminary numerical results are encouraging because they
predict halo losses which seem under control. Indeed the $HIDIF$
scenario gives a total loss of beam power per meter which is about a
thousand higher than the $LHC$; however in both cases the estimated 
losses appear much smaller than the permissible $1$ Watt/m.

\end{document}